\documentclass[preprint, 3p, twocolumn]{elsarticle}
\usepackage{graphicx}
\usepackage{booktabs}
\usepackage{tabularx}
\usepackage[normalem]{ulem}
 \useunder{\uline}{\ul}{}
 \usepackage{rotating}
 \usepackage{dsfont}
\usepackage{color}
\def\Manu#1{\textcolor{black}{#1}}

\usepackage[switch]{lineno}
\usepackage{graphicx,xcolor}
\usepackage{epstopdf}
\usepackage{setspace}
\usepackage[T1]{fontenc}
\usepackage{amssymb,amsfonts,amsmath}
\usepackage{setspace}
\usepackage{subcaption}
\usepackage{mathtools}

\begin{document}

\title{\Manu{Hyperspectral imaging for dynamic thin film interferometry}}
\author[vinny]{V. Chandran Suja}
\ead{vinny@stanford.com}
\author[johnny]{J. Sentmanat}
\author[greg]{G. Hofmann}
\author[greg]{C. Scales}
\author[vinny]{G.G. Fuller}\ead{ggf@stanford.com}
\address[vinny]{Department of Chemical Engineering, Stanford University, Stanford, California 94305}
\address[johnny]{Department of Mechanical Engineering, Texas A\&M University, College Station, Texas 77843}
\address[greg]{Johnson \& Johnson Vision Care Inc., Jacksonville, FL 32256, United States}

\begin{abstract}
Dynamic thin film interferometry is a technique used to non-invasively characterize the thickness of thin liquid films. Recovering the underlying thickness from the captured interferograms, unconditionally and automatically is still an open problem. Here we report a compact setup employing a snapshot hyperspectral camera and the related algorithms for the automated determination of thickness profiles of dynamic thin liquid films. The proposed technique is shown to recover film thickness profiles to within $100\;nm$ of accuracy as compared to those profiles reconstructed through the manual color matching process. Subsequently, we discuss the characteristics and advantages of hyperspectral interferometry including the increased robustness against imagining noise as well as the ability to perform thickness reconstruction without considering the absolute light intensity information. 
\end{abstract}


\maketitle

\section{Introduction}




Hyperspectral imaging is a spectral imaging technique that combines spectroscopy and digital photography,  thus yielding a spectrum at each pixel in the image of a scene.  Due to the enhanced spectral resolution, hyperspectral imaging has been routinely used for many remote sensing applications such as monitoring agriculture and vegetation, and for detecting mineral and oil deposits \cite{wang2016hyperspectral,lawrence2003calibration,singh2009detection}. Recently, hyperspectral imaging techniques have proliferated medical imaging \cite{lu2014medical}, including for cancer tissue detection \cite{kong2006hyperspectral, dicker2006differentiation} and ophthalmology \cite{johnson2007snapshot,li2017snapshot}. Furthermore, hyperspectral imaging is increasingly being used for industrial machine vision applications such as for food and process quality control \cite{ liu2014recent, sun2010hyperspectral}. As hyperspectral imagers are rapidly becoming more robust, compact and economical, this technology is poised to significantly revolutionize machine vision in many different settings. 

One such setting is dynamic thin film interferometry \cite{frostad2016dynamic}. Thin film interferometry is a technique that employs light interference to measure thickness of films that are usually a few microns thick. Thin film interferometry has been used since the early 1950's to characterize static films \cite{scott1950thickness, blodgett1952step} and in the subsequent years the technique was adopted for studying dynamic thin liquid films \cite{sheludko1967thin}. This technique has since then become the conventional method for characterizing thin liquid films in research settings, primarily due to its simplicity and non-invasive nature. Some of the important areas that make use of this technique includes characterizing foams \cite{pugh2005experimental, denkov2004mechanisms} and emulsion  \cite{politova2017factors} stability, drop impacts \cite{tran2012drop}, and film coatings \cite{doane1998tear, bhamla2016instability}. Despite the popularity of the technique in research settings, dynamic thin liquid film interferometry has not proliferated industrial or commercial settings.    

The primary obstacle that prevents the wide spread commercial use of interferometry for characterizing thin films is the difficulty in automatically analyzing the interferograms to recover the underlying film thicknesses.  The automatic analysis is complicated by the transcendental phase-periodic governing equations that non-uniquely relates the pixel intensities in interferograms to film thicknesses. This inherent complexity is amplified by the uncertainty introduced by the unavoidable imaging and background noise. As a result, the unconditional determination of film thickness from interferograms has remained an unsolved research problem since the inception of this technique. 

\begin{figure*}[!h]
\centering
\includegraphics[width=0.99\linewidth]{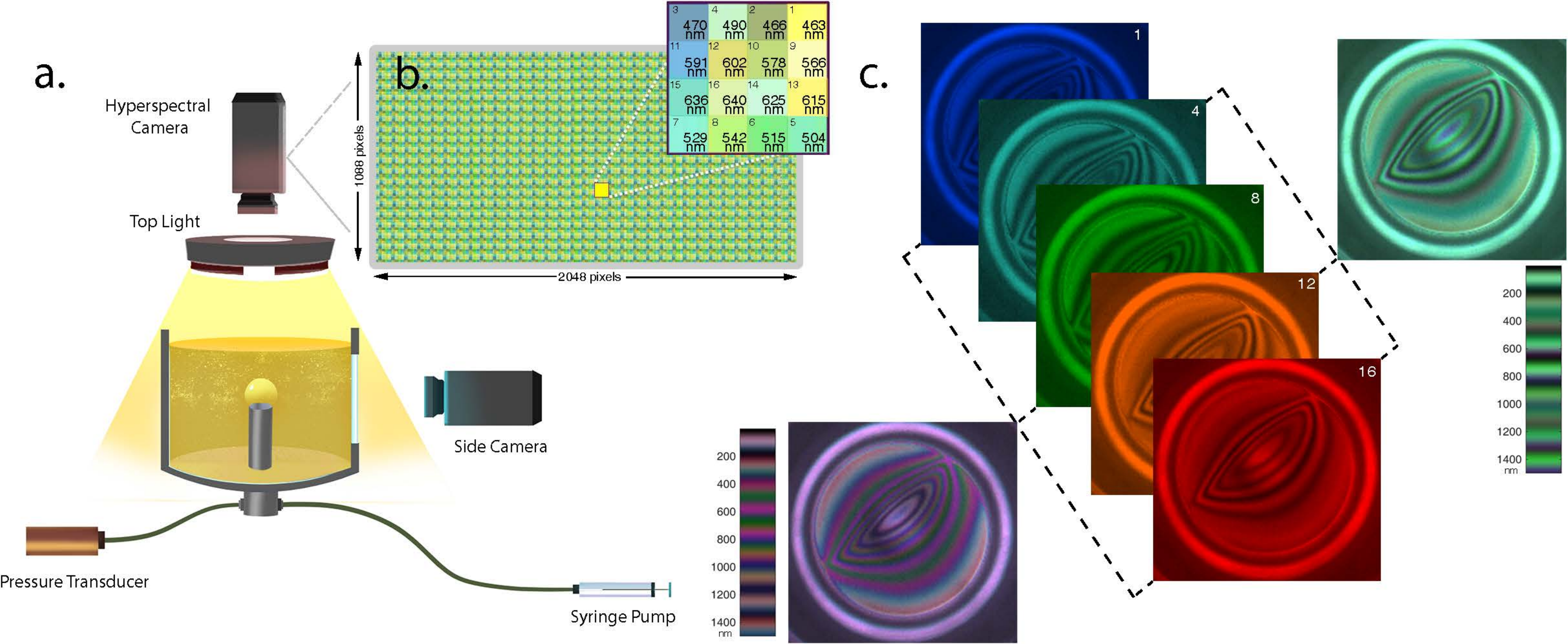} \caption{Schematic of the compact experimental setup along with the details of hyperspectral camera used in this study. {\bf a.} The experimental arrangement used to record the hyperspectral interferograms. Further details are available in previously published works \cite{frostad2016dynamic}. {\bf b.} Details of the fabry-perot filter array inside the snapshot hyperspectral camera. Each fabry-perot filter in the repeating filter array unit is numbered according to the ascending order of the peak wavelengths of the filters in that unit. {\bf c.} Five slices from the HSI cube along with two RGB composites generated by combining  bands 1,8,16 and bands 4,12,16. Such RGB composites are useful for visualizing hyperspectral interferograms, and in this case also qualitatively illustrates how hyperspectral imaging can overcome the non-uniqueness between film thickness and color associated with the traditional RGB interferometry.  }\label{fig.ExperimentalSetup}
\end{figure*}

Researchers have traditionally side stepped this problem broadly by the following approaches. One of the approaches involves manually identifying the film thickness at a region utilizing a reference colormap \cite{frostad2016dynamic, goto2003computer}.  This approach is quite robust, however it is quite slow and suffers from human subjectivity. Another common approach utilizes fringe counting from a known absolute reference thickness in the film \cite{sett2013gravitational, zhang2015domain, bluteau2017water}. This approach is quite fast but is not robust and requires assumptions on the spatial structure of the film. Yet another approach involves pre-calibrating the pixel intensities in interferograms to film thicknesses utilizing liquid films of known thickness profiles \cite{gustafsson1994measuring, hartl1999automatic}. This approach is quite fast and robust, however this technique is restricted to scenarios where such calibrations are possible.  

In this manuscript we report an arrangement utilizing hyperspectral imaging for the unconditional and automated reconstruction of thickness profiles from thin liquid film interferograms. Approaches utilizing hyperspectral imaging (using the pushbroom like techniques) have previously been reported for characterizing static thin films \cite{cabib1999film, tremmel2017inline}. However, these approaches are not suited for dynamic films,  primarily due to the nature of image acquisition, and to the best of our knowledge have never been modified for use with dynamic films. Here, we describe a compact setup employing a snapshot hyperspectral imager and the related algorithms for automated determination of thickness profiles of thin films. The accuracy of this technique is established by comparing the absolute pixel-wise differences across the manually reconstructed and the automatically reconstructed thickness profiles. Finally, the key advantages and unique characteristics of hyperspectral thin film interferometry are also discussed.

\section{Methods}

\subsection{Theory}
The theory of thin film interference was formalized in the early 19\textsuperscript{th} century by Fresnel, and has since been discussed by many researchers in the context of measuring the thickness of thin films such as for bubbles \cite{sheludko1967thin}, tear films \cite{doane1989instrument}, and for surface profiling \cite{blodgett1952step}. Here we will briefly develop a formulation relevant for a hyperspectral camera.

Consider a beam of light having intensity $I_0(\lambda)$ incident on a thin liquid film of thickness $d$ and refractive index $n_2$. The film is bounded on top and bottom by media having refractive indices $n_1$ and $n_3$ respectively. Assuming normal incidence and non dispersive films, the reflected light intensity $I(d,\lambda)$ emanating from the thin film can be written as, 

\begin{align}\nonumber
    I(\lambda,d) &= I_0(\lambda) \left( R_1 +R_2(1 - R_1)^2  \right.\\ 
    &\left. + 2\sqrt{R_1R_2(1 - R_1)^2}\cos\left( \frac{4\pi n_2 d}{\lambda} \right)\right)
\end{align}
Here $\lambda$ is the wavelength of light, while $R_1$ and $R_2$ are the reflectivity coefficients obtained from the Fresnel equations evaluated for normal incidence, and are given by,
\begin{align}
    R_1 = \left(\frac{n_1 - n_2}{n_1 + n_2}\right)^2\\
    R_2 = \left(\frac{n_2 - n_3}{n_2 + n_3}\right)^2
\end{align}
Finally, the intensity perceived by the $i^{th}$ channel of a pixel $H$ in a hyperspectral camera as a function of the film thickness can be computed as, 
\begin{equation}\label{eq.HyperspectralIntensities}
    H_i (d) = \int_{\lambda_0}^{\lambda^f}  I(\lambda,d) I_r(\lambda) S_i(\lambda)
\end{equation}
Here  $I_r(\lambda)$ is the spectral response of filters in the system, and $S_i(\lambda)$ is the spectral sensitivity of the $i^{th}$ channel of a pixel. 

During an experiment (Fig.\ref{fig.ExperimentalSetup}), a hyperspectral camera having $h$ channels at every pixel will encode reflections from a thin film of thickness $d$ as a $h$ dimensional vector. Utilizing Eq.\ref{eq.HyperspectralIntensities}, we can invert this $h$ dimensional vector to recover the thickness of the thin film. In section \ref{sec:Results}, we will show that using a hyperspectral camera for this process instead of an RGB camera has significant advantages such as increased robustness against noise and absolute light intensity independent thickness determination.    

\subsection{Experimental Setup}
The single bubble coalescence experiments used to validate the utility of hyperspectral imaging for thin film thickness measurements were performed using a modified Dynamic Fluid-Film Interferometer (DFI). The construction \cite{frostad2016dynamic} and the utility \cite{suja2018evaporation,kannan2018monoclonal,jaensson2018tensiometry} of the DFI has been previously discussed in a number of publications and in references therein. 

For the current study, the DFI was modified to have a 16 channel snapshot HyperSpectral Imaging (HSI) camera (Model: MQ022HG-IM-SM4X4-VIS, Manufacturer: Ximea GmbH, Germany) as its top camera (Fig.\ref{fig.ExperimentalSetup}a). As the filter array inside the HSI camera has narrow spectral response bands (\textit{\textbf{Supplementary Figure S4}}), the dichroic triband filter utilized with the top light for the same purpose (reducing the FWHM of spectral bands of a RGB camera)\cite{frostad2016dynamic} was removed.  The removal of the dichroic filter thus resulted in 120\% increase in the luminous flux entering camera - improving the signal to noise ratio in the acquired hyperspectral interferograms. 

To benchmark the thin film measurement capability of the hyperspectral camera, single bubble experiments were also performed using RGB cameras (IDS UI 3060CP), commonly used for thin film interferometery \cite{frostad2016dynamic,suja2018evaporation}.

\subsection{Image processing}
\begin{figure}[!h]
\centering
\includegraphics[width=\linewidth]{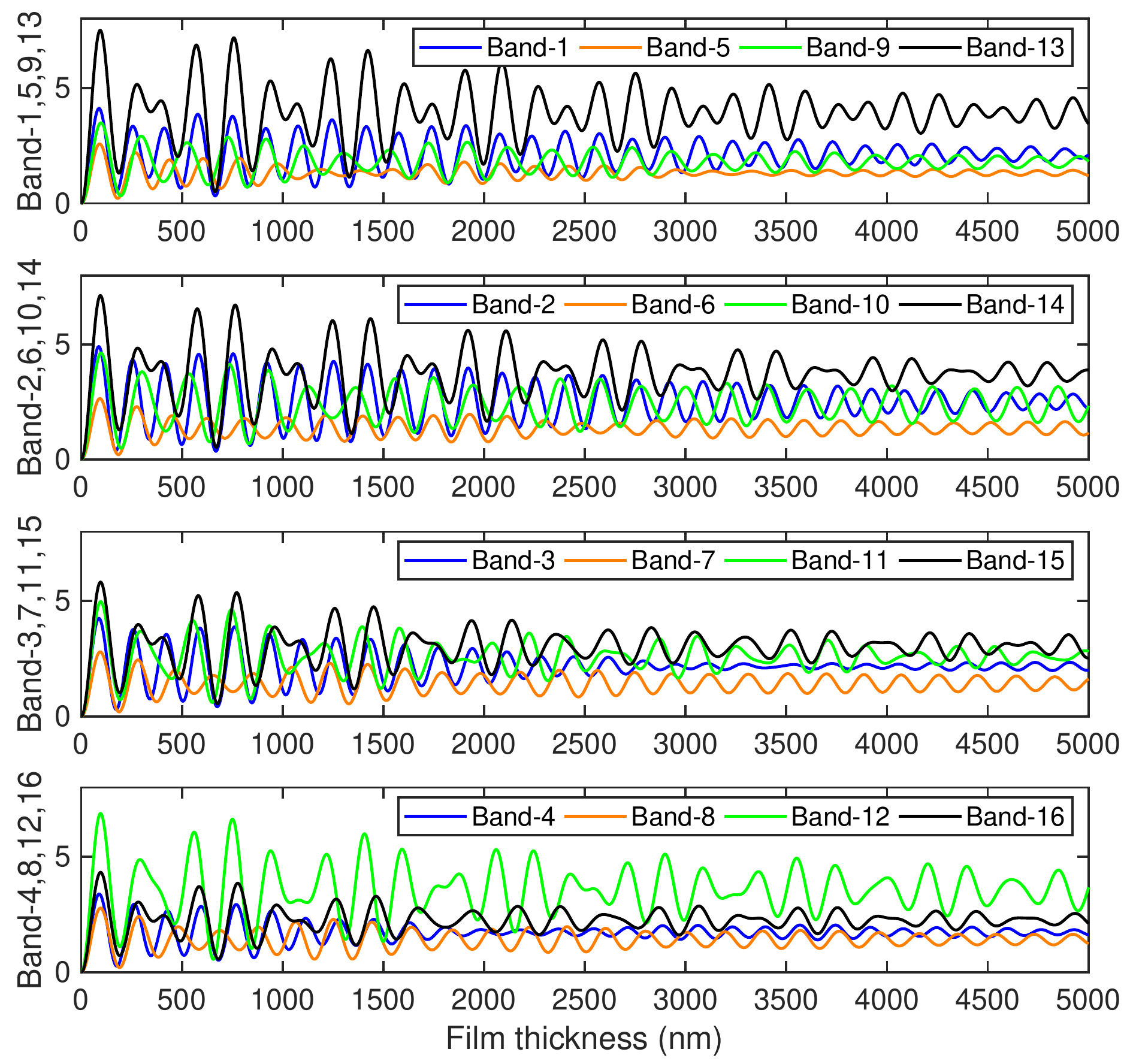} \caption{The spectral map showing the theoretical spectral signature perceived by the different channels in the camera as a function of the film thickness. Here we have assumed a silicone oil film bounded by air on either side (refractive index triplet - [1 1.4 1]). The spectral response data of the camera is available in \textit{\textbf{Supplementary Figure S4}}. Note that there is a continuous variation of the spectrum as a function of the film thickness.} \label{eq.SpectrumMap}
\end{figure}

To recover the film thickness from the hyperspectral image, the following steps were executed utilizing Matlab.
Initially, the raw images from the snapshot HSI camera were sliced and spliced appropriately to reconstruct the hyperspectral cubes. Subsequently, background subtraction was performed on the cubes, followed by cropping, intensity correction and normalization. A k-Nearest neighbour search utilizing the cosine distance metric is performed between each pixel in the resulting HSI cube and the theoretical spectral map (Fig.\ref{eq.SpectrumMap}) generated from Eq.\ref{eq.HyperspectralIntensities}. The thicknesses obtained as the first nearest neighbour in the k-Nearest neighbour search is used to construct the initial estimate of the thickness profile.   

Finally, a spatial optimization algorithm is utilized to correct for the any incorrectly assigned points. The algorithm basically enforces the $C^0$ spatial continuity in film thickness by replacing any incorrectly assigned thickness with an appropriate thickness from the k possible thickness values at that point.   

\begin{figure*}[!h]
\centering
\includegraphics[width=\linewidth]{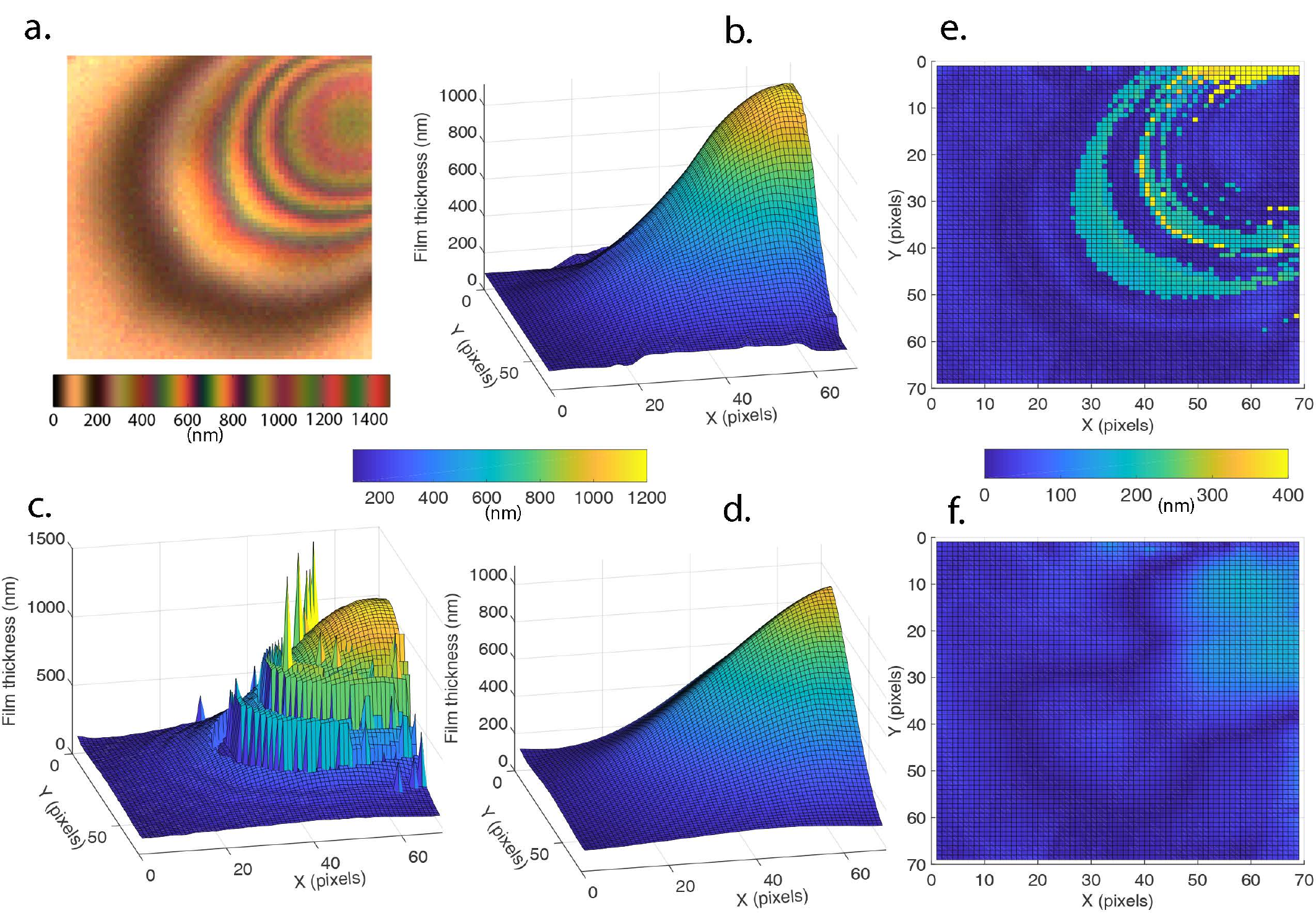} \caption{The thickness reconstruction performance of the proposed snapshot hyperspectral interferometry technique. {\bf a.} The RGB visualization of the hyperspectral interferogram (and its colormap) that is used to illustrate the reconstruction performance. {\bf b.} The manually reconstructed thickness profile\cite{frostad2016dynamic}  (used here as the ground truth) {\bf c.} The corresponding automatically reconstructed thickness profile. {\bf d.} The automatically reconstructed thickness profile after optimization. {\bf e.} The absolute pixel wise height difference between the manual and the unoptimized thickness profile. Over $80\%$ of the pixels differ by $100\; nm$ or less. {\bf f.} The absolute pixel wise height difference between the manual and the optimized thickness profile. Over $90\%$ of the pixels differ by $100\; nm$ or less. }\label{fig.HSIReconstruction}
\end{figure*}

\section{ Results and Discussion}\label{sec:Results}
\subsection{Thickness reconstruction performance}

The performance of the new camera system is established by reconstructing the dynamic film thickness of a bubble in a silicone oil mixture. The results before and after optimization are compared to the manually reconstructed thickness profiles \cite{frostad2016dynamic} in Fig.\ref{fig.HSIReconstruction}. The reconstructed thickness profile before optimization (Fig.\ref{fig.HSIReconstruction}.c) broadly resembles the manually reconstructed profile (Fig.\ref{fig.HSIReconstruction}.b). However, due to noise and spectral mixing in the interference data, there are regions in the reconstructed profile that have physically unrealistic gradients. These artifacts are removed utilizing the optimization routines, and the resultant thickness profile after optimization (Fig.\ref{fig.HSIReconstruction}.d) is seen be very similar to the manually reconstructed profile. The mean film thickness ($\iint T(x,y) dx dy /\iint dx dy  $), a very common metric used to report the thickness of thin films \cite{frostad2016dynamic,suja2018evaporation,suja2020foam}, is very similar across the three cases with values of $358.9\;nm$, $344.4\;nm$ and $347.9\;nm$ for the manual, unoptimized and optimized thickness profiles respectively.

\begin{figure*}[!t]
 \centering
 \subcaptionbox{}
 {\includegraphics[width=0.5\textwidth]{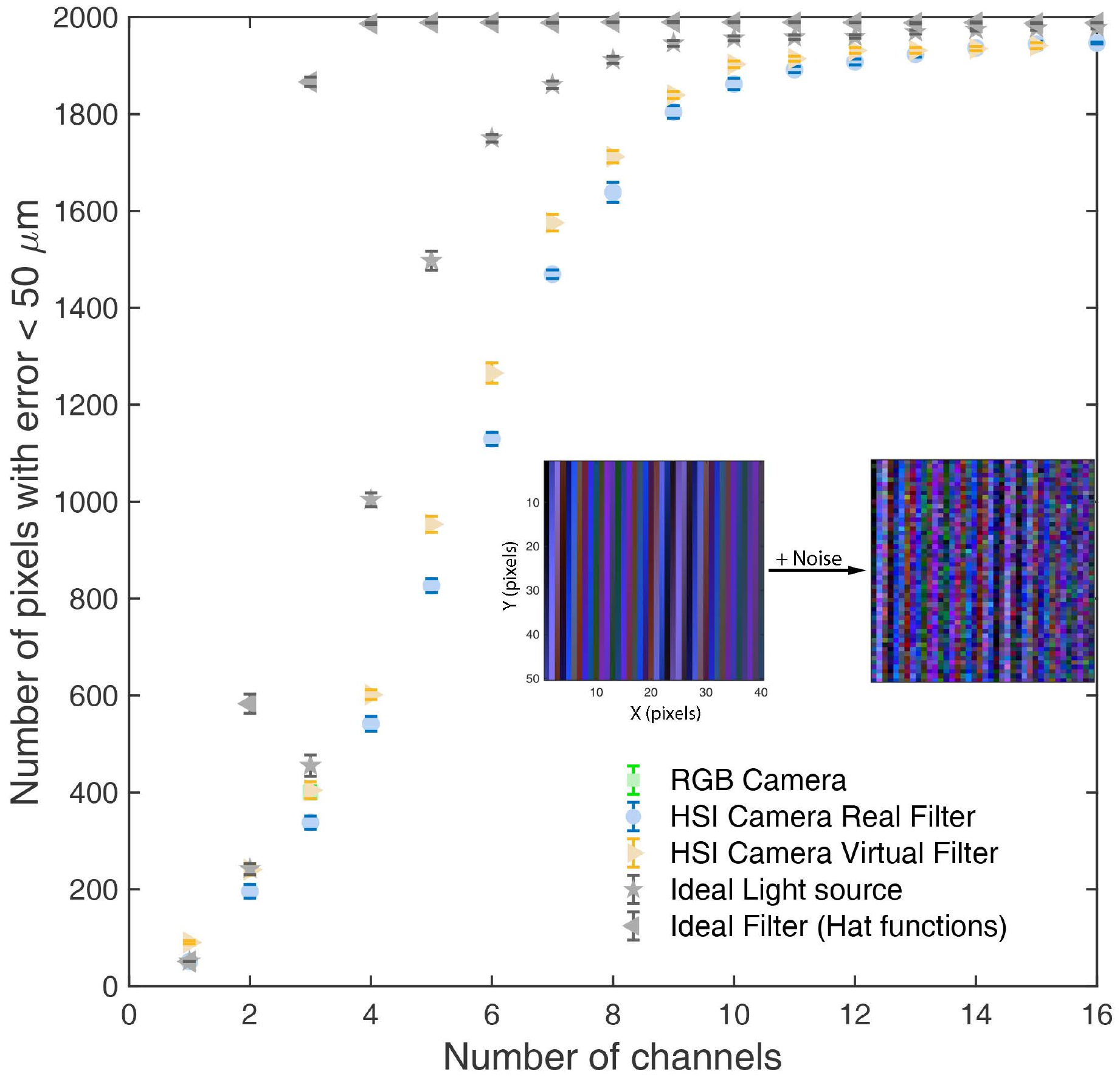}}\subcaptionbox{}{\includegraphics[width=0.5\textwidth]{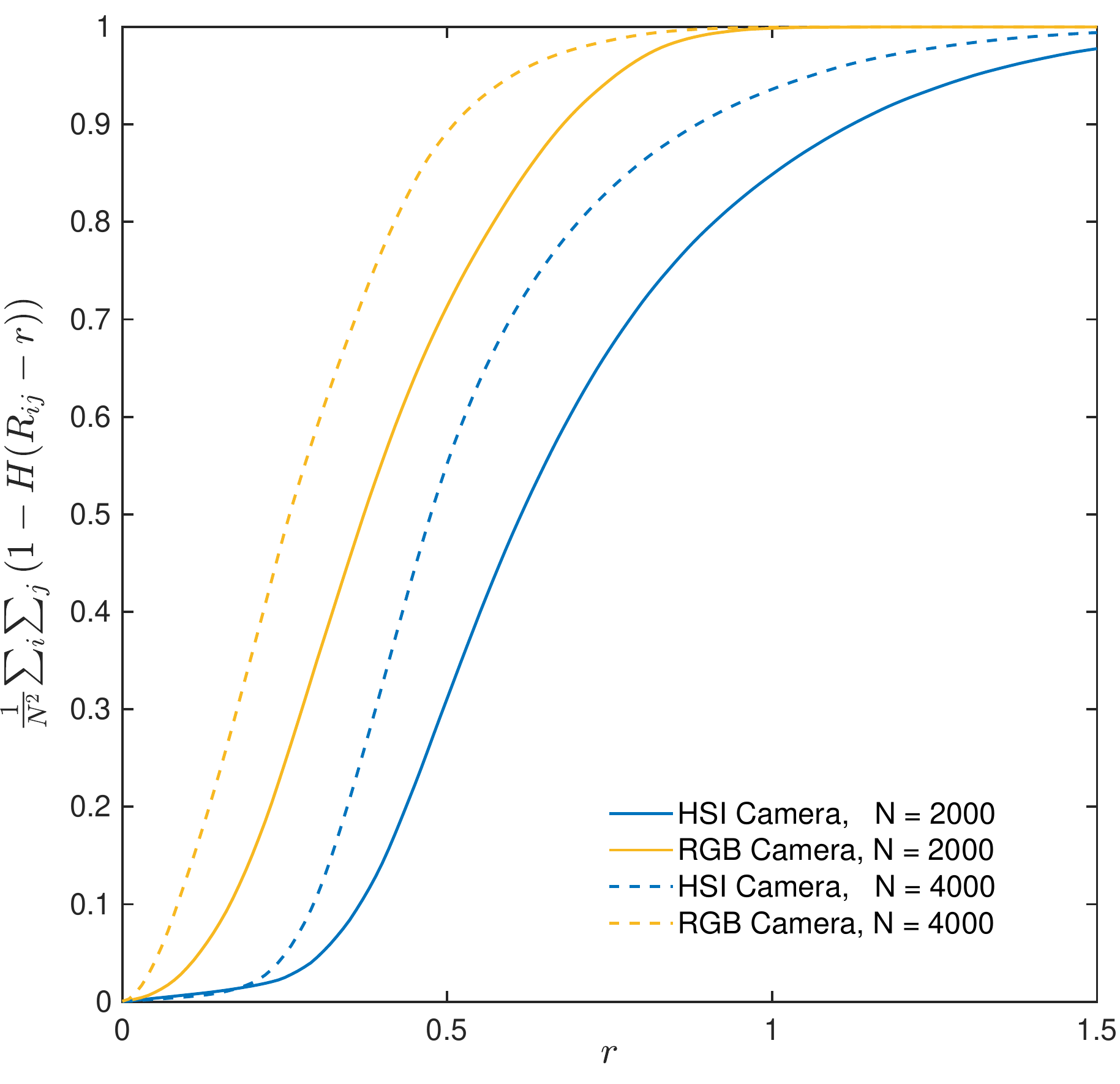}}
  \caption{(a) Robustness against noise as a function of number and type filters in a camera. The reconstruction accuracy is seen to increase with the number of channels in the camera. Further, a hypothetical camera with perfect (hat function) filters is seen to require to the least number of channels for a high fidelity thickness reconstruction. (b) The cumulative pair wise distance distributions for the theoretical colormap  (RGB camera) and spectralmap (HSI camera) entries. Here N is the number of contiguous thickness values in the theoretical maps starting at $0\; nm$ and separated by $1\;nm$. The spectral entries corresponding to a hyperspectral camera are separated by a larger distance ($r$) as compared to that of a RGB camera.}\label{fig:RobustnessAgainstNoise}
  \label{fig:expPressure}
\end{figure*}

The absolute pixel-wise height difference between the manually reconstructed and the automatically reconstructed thickness profiles pre and post optimization are shown in Fig.\ref{fig.HSIReconstruction}.e,f. Clearly, the reconstructions are very accurate with over $80\%$ of the pixels differing by less than $100 \;nm$ for the unoptimized case, while over $90\%$ of the pixels differ by less than $100 \;nm$ for the optimized case. The remaining $10\%$ of the regions have errors larger than $100\;nm$, and are most likely a result of inadequate background subtraction as well as due to the lac of pixel by pixel calibration information for the camera.  As a reference, its also worth noting that reconstructions using the 3 channel data available from a RGB camera has an  unacceptable accuracy, with less than $20\%$ of the pixels being classified within $100 \;nm$ (\textit{\textbf{Supplementary Figure S3}}).

Finally, a convenient advantage of the automated reconstruction is the ease of analyzing time sequential data to obtain the temporal evolution of the film thicknesses. \textit{\textbf{Supplementary Video V1}} shows one such reconstruction of a temporally evolving film thickness profile alongside its corresponding raw interferogram.

\subsection{Robustness against noise}
Theoretically, the 3 channels in RGB interferograms are sufficient to disambiguate the underlying film thicknesses \cite{gustafsson1994measuring}. However in practice, the inherent noise in the acquired image data breaks this theoretical uniqueness between the RGB intensities and the film thickness; thus requiring optimization (even for static films) \cite{kitagawa2013thin} and/or calibration \cite{gustafsson1994measuring} for thickness recovery. Unlike the RGB interferograms, hyperspectral interferograms are relatively more robust to noise.  

To illustrate the robustness of hyperspectral interferograms against noise, we construct a ramp thickness profile having a minimum thickness of $1\; nm$ and a maximum thickness of $2000\;nm$ (\textit{\textbf{Supplementary Figure S1}}). This thickness profile is mapped to corresponding color interferograms using the color maps of the tested cameras. Subsequently, Gaussian noise (see \textit{\textbf{Supplementary Text}} is added to the images. Inset in fig.\ref{fig:RobustnessAgainstNoise}(a) shows one such RGB image before and after addition of the noise. Reconstruction algorithms (without optimization) are then used to predict the thickness ($T$) at each pixel in the noisy images. $T$ can be compared to the ground truth thickness ($T^g$) to quantify the reconstruction accuracy. We visualize the reconstruction accuracy in fig.\ref{fig:RobustnessAgainstNoise}(a) by plotting the number of pixels having an error of less than $50 \mu m$ $\left( \sum_i  \mathds{1}(|T^g_i - T_i|<50)\right)$ as a function of the number of channels in the camera.

From fig.\ref{fig:RobustnessAgainstNoise}(a) we see that for an RGB camera only about 20\% of the pixels are classified within acceptable error. However, for a HSI camera (utilizing all 16 channels) we are able to recover the thickness from 97\% of the pixels within acceptable error; thus showing that HSI cameras are more robust to noise. For the tested HSI camera, we also obtain similar reconstruction accuracy using either the real or virtual filters (\textit{\textbf{Supplementary Figure S5}}) on a HSI camera, thus suggesting any type of filter may be used in the reconstruction process. Further we also see that having an ideal light (uniform intensity across wavelengths) or ideal filters (responses given by hat functions) can enhance the reconstruction performance; with ideal filters having the greatest impact on reconstruction accuracy (see \textit{\textbf{Supplementary Text}}). Finally, its also interesting to note that the results indicate that a 3 channel camera (RGB camera) with perfect filters has a performance comparable to a regular HSI camera utilizing all 16 channels. 

The increased robustness of HSI interferograms against noise is related to the higher dimensionality of the color co-ordinates that correspond to a given film thickness. As a result of the higher dimensionality, the pair wise Eucledian distances of HSI color co-ordinates are larger than the corresponding pair wise distances of color co-ordinates obtained from a RGB camera. This observation is quantified in fig.\ref{fig:RobustnessAgainstNoise}(b) by plotting the cumulative pair wise distance distribution of the color co-ordinates (resolved at every nanometer) for a HSI and a RGB camera for two different film thickness ranges - $[0,2000]$ and $[0,4000]\;nm$. From fig.\ref{fig:RobustnessAgainstNoise}(b), we clearly see that for a given thickness range, the curves corresponding to HSI cameras as compared to RGB cameras are shifted towards larger distances ($r$). Hence, as a consequence of the increased separation of HSI color co-ordinates, perturbations due to noise are less likely to result in a color co-ordinate becoming similar to another.



\subsection{Absolute Light intensity independent thickness reconstruction}
Another advantage of using hyperspectral imaging is that the absolute intensity of light at any point in the interferogram may be neglected during thickness reconstruction. This can shown be true using two different arguments. Firstly, standard transformations (such as RGB to HSV) can isolate the intensity information into a single channel. Neglecting the intensity channel, still gives us (for the tested HSI camera) information from 15 channels to unambiguously (see Fig.\ref{fig:RobustnessAgainstNoise}) reconstruct the thickness. Since an RGB camera has only 3 channels, neglecting the intensity makes it impossible to unambiguously determine the thickness. Secondly, the cosine distance metric utilized in this paper to determine the film thicknesses completely ignores the intensity information. Despite ignoring the intensity information, we are able to recover the thickness profiles quite well (Fig.\ref{fig.HSIReconstruction}.d), thus confirming that hyperspectral imaging enables thickness reconstruction without necessarily requiring information about the absolute light intensity. 

A direct consequence of this result is that the reconstruction techniques described in this paper are robust against the natural spatial variation of incident light intensity (vignetting) over the interferogram. Hence in addition to obviating the need for accurately obtaining the absolute light intensities, the corrections for vignetting may also be conveniently avoided. Note that flat field corrections may still be required if there are spatial variations in pixel sensitivities in a camera. 

\subsection{Characteristics of Hyperspectral thin film interferometry}
Hyperspectral imaging when applied to thin film interferometry has some interesting characteristics. Firstly, the number of spectral classes is higher than in traditional hyperspectral imaging used for remote sensing or medical imaging. The number of spectral classes in thin film interferometry goes as $\mathcal{O}(T_{max}/T_{resolution})$, which is equal to $4000$ when trying to resolve film thickness of upto $ T_{max} = 4000\;nm$ with a resolution of $ T_{resolution} = 1 \;nm$. Consequently, spectral matching routines used for thin film interferometry should have a high degree of specificity, and may also need optimization algorithms (as used in this paper) to completely reconstruct the thickness profile.   

Secondly, the spectra vary gradually across spatially adjacent spectral classes.  Unlike traditional hyperspectral imaging where there are no restrictions on spatially adjacent classes (and hence on the spectra), the spatial continuity of thin films restricts the spatially adjacent classes to correspond to contiguous film thicknesses. Hence, as a consequence of Fig.\ref{eq.SpectrumMap}, there is a gradual variation of the spectrum across spatially adjacent classes. Consequently, spectral mixing (due to lack of sufficient camera resolution) does not pose difficulties during thickness reconstruction, thereby obviating any need for spectral unmixing techniques \cite{villa2011spectral}.  

Thirdly, as thin liquid films are dynamic, the spectral signatures change rapidly both in space and time. As a consequence, snapshot hyperspectral imaging (as opposed to techniques such as the pushbroom) is better suited for thin film interferometry.

\section{Conclusions}
In this paper we reported a compact setup employing snapshot hyperspectral imaging and the related algorithms for the automated determination of thickness profiles of dynamic thin liquid films. We showed that we can reconstruct dynamic film profiles to within $100 \; nm$ of those reconstructed manually. As manually reconstructed profiles themselves are only accurate upto $50 \; nm$, the automatic reconstructed profiles are a faithful representation of the ground truth. We also showed that hyperspectral interferometry has two key advantages, namely, the increased robustness against imaging noise and the ability to neglect the absolute light intensity information during thickness reconstruction.     

Future studies may be focused on improving both the hardware and software aspects presented in this work. Improvements in the spatial resolution, the sensor quantum efficiency, and the filter properties like transmissivity and FWHM of the hyperspectral imager will enhance the reconstruction performance of the system. Our results also suggest that improving the filters (especially the Q-factor) will have a significant impact on the robustness of the system. Finally improvements in image pre-processing (like background substraction) and in spectral matching algorithms (such as enhancing the specificity and robustness) will also aid in perfecting the proposed system.                                                                                         
\section*{Acknowledgements}
We thank Akshaya Baskar from Johnson \& Johnson Vision Care Inc. (JJVCI) for her help in deploying and testing the setup at JJVCI, Kevin Toerne from Ximea corp for his enthusiastic assistance with setting up the hyperspectral imaging hardware, and Prem Sai for creating the schematic illustrations in the manuscript. This study was supported by a grant from JJVCI (Grant no. SPO 132650).

\section*{References}
\bibliographystyle{vancouver}
\bibliography{Reference}

\end{document}